# The Carnegie Institution Geophysical Laboratory Seminar,
# Analysis of evidence of Mars life
held  05/14/2007

5251 Broad Branch Avenue NW, Washington, DC 20015; Main Phone: 202-478-8900; Fax: 202-478-8901

Summary of the lecture given by

**Gilbert V. Levin, Ph.D.**

Chairman, Executive Officer for Science, Spherix Incorporated, Beltsville, MD 21705, US; phone 301-419-3900; fax 301-210-4908; e-mail: *glevin@spherix.com*

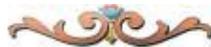

**Abstract**:  *Gillevinia straata*, the scientific name [1, 2] recognizing the first extraterrestrial living form ever nomenclated, as well as the existence of a new biological kingdom, Jakobia, in a new biosphere -Marciana- of what now has become the living system Solaria, is grounded on old evidence reinterpreted in the light of newly acquired facts. The present exposition provides a summary overview of all these grounds, outlined here as follows. A more detailed paper is being prepared for publication.

__________

In a May 3 Carnegie dinner, Carnegie Institution Chairman, Michael Gellert, pointed out that the Institution was founded to - and does - concentrate on high risk problems. This makes Carnegie the proper venue for exploring a major scientific paradigm change - there is life on Mars. And, most importantly, to determine whether life had more than one origin, as would be indicated were Earth life and Mars life fundamentally different. Such a result would have profound implications for the existence of life, including intelligent life, throughout the universe. I am thus very pleased to have the opportunity to present this prospect at the Carnegie Institution Geophysical Laboratory seminar.

1. The Viking landers carried nine courses of the Labeled Release experiment (LR) designed to detect any metabolizing microorganisms that might be present on the martian surface. The LR was designed to drop a nutrient solution of organic compounds labeled with radioactive carbon atoms into a soil sample taken from the surface of Mars and placed into a small test cell. A radiation detector then monitored over time for the evolution of radioactive gas from the sample as evidence of metabolism: namely, if microorganisms were metabolizing the nutrients they had been given. When the experiment was conducted on both Viking landers, it gave positive results almost immediately. The protocol called for a control in the event of a positive response. Accordingly, duplicate soil samples were inserted into fresh cells, heated for three hours at 160 ºC to sterilize them (the control procedure established for all Viking biology experiments), allowed to cool and then tested. These courses produced virtually no response, thus completing the pre-mission criteria for the detection of microbial life. All LR results support, or are consistent with, the presence of living microorganisms. Yet between 1976 and late 2006 life on Mars remained a subject of debate, with the scientific consensus being negative because of the following arguments:

   a. The Viking organic analysis instrument (GCMS), an abbreviated gas chromatograph-mass spectrometer designed to identify the organic material widely presumed to be present on Mars, found no organic molecules. After years of discussion and experimentation, a consensus was reached explaining this negative result as a lack of sensitivity [see 3].

b. "UV destroys life and organics". Yet sampling soil from under a rock on Mars demonstrated that UV light was not inducing the LR activity detected.

c. "Strong oxidants were present that destroy life and organics". Findings [4] by the Viking Magnetic Properties Experiment showed that the surface material of Mars contains a large magnetic component, evidence against a highly oxidizing condition. Further, three Earth-based IR observations, by the ESA orbiter [5] failed to detect the putative oxidant in any amount that could cause the LR results, and, most recently, data from the Rover Opportunity have shown Mars surface iron to be not completely oxidized (ferric) - but to occur mostly in the ferrous form which would not be expected in a highly oxidizing environment.

d. "Too much too soon". The LR positive responses and their reaction kinetics were said to be those of a first order reaction, without the lag or exponential phases seen in classic microbial growth curves, all of which seemed to argue for a simple chemical reaction. However, terrestrial LR experiments on a variety of soils produced response rates with the kinetics and the range of amplitudes of the LR on Mars, thereby offsetting this argument.

e. Lack of a new surge of gas upon injection of fresh medium. Although 2nd injection responsiveness was not part of the LR life detection criteria, the lack of a new surge of gas upon injection of fresh medium on an active sample was interpreted as evidence against biology. However, a previous test of bonded, NASA-supplied Antarctic soil, No. 664, containing less than 10 viable cells/g [6], had shown this same type of response to a 2nd injection. The failure of the 2nd injection to elicit a response can be attributed to the organisms in the active sample having died sometime after the 1st injection, during the latter part of Cycle 1. The ef-fect of the 2nd injection was to wet the soil, causing it to absorb headspace gas. The gradual re-emergence of the gas into the headspace with time occurred as the system came to equilibrium.

f. "There can be no liquid water on the surface of Mars". Since November and December 2006, the accumulated evidence shows that liquid water exists in soil even if only as a thin film. Viking, itself, gave strong evidence [7] of the presence of liquid water when the rise in the temperature of its footpad, responding to the rising sun, halted at 273 degrees K. Snow or frost is seen in Viking images of the landing site (*e.g.*, Viking Lander Image 21I093). Pathfinder has shown that the surface atmosphere of Mars exceeds 20 °C part of the day, providing transient conditions for liquid water. Together, these observations constitute strong evidence for the diurnal presence of liquid water. In explaining the stickiness of the soil, MER scientists have said that it "might contain tiny globules of liquid water," or "might contain brine". Other images of Mars show current, if intermittent, rivulet activity. On the Earth's South Polar Cap and within permafrost in the Arctic there is liquid water: even in those frozen places, very thin films of liquid water exist among the interstices of ice and minerals, enough to sustain an ecology involving highly differentiated species.

g. "Cosmic radiation destroys life on Mars". a recent report [8] calculated the incoming flow of both galactic cosmic rays particles (GCR) and solar energetic protons (SEP) over a wide energy range. As a result one may acknowledge that -without even invoking natural selection to enhance radiation protection and damage repair- the radiation incident to the surface of Mars appears trivial for the survival of numerous terrestrial-like microorganisms. With respect to the near-term effect of the radiation, when Surveyor's camera was returned from the Moon after being in its much-harsher-than-Mars radiation fields for forty months, it was found to contain viable microorganisms. However, the point was then made that exposures of constantly frozen microorganisms to this flux for millions to billions of years would have damaged their DNA and its repair mechanism to the point where survival could not occur. In this regard, Viking and the Pathfinder thermal data demonstrate that, at least at the three widely separated locations of those landers, prolonged freezing is not the case.

2. Those arguments should have been satisfied with the cited data. If not, additional evidence added an even richer context in support of the LR results. Main items are listed as follows.

3. Further supporting evidence includes the possible presence, on some of the Martian rocks, of desert varnish, a coating which on Earth is of microbial origin or contains products generated by microorganisms - an observation originally made by Viking on which several recent articles have rekindled interest. Adding to this rising tide of facts supporting the detection of life by the Viking LR experiment are the recent findings in the Martian atmosphere of methane, formaldehyde, and, possibly, ammonia, gases frequently involved in microbial metabolism. The existence of the short half-lived, UV-labile methane requires a source of continual replacement. Continual volcanic activity, a potential non-biological source of methane, has not been indicated by thermal mapping of the entire planet. In the Earth's atmosphere, methane is sustained primarily by biological metabolism. Moreover, the methane detected on Mars was associated with water vapor in the lower atmosphere, consistent with, if not indicative of, extant life.

4. As still further evidence, the kinetics of evolution of labeled gas in the Viking LR experiment indicates the possibility of a circadian rhythm, daily over the length of the experiments, up to 90 sols. However, as of now, these are only indications, not statistically significant, as is pointed out in two papers of which I'm a coauthor [9, 10]. However, another paper [11], using a non-linear approach, con-cluded, "Our results strongly support the hypothesis of a biologic origin of the gas collected by the LR experiment from the Martian soil." A new study, in which the authors of the initial papers and the most recent paper are collaborating, is currently underway to further investigate the statistical significance for that conclusion.

5. Huge recent advances in the research of the variety of extremophiles on Earth have added very strong import to the current context. Recently, an expert in soil science from the Netherlands communicated to the congress of the European Geosciences Union that the discovery of the recent detection of phyllosilicate clays on Mars may indicate pedogenesis processes, or soil (as opposed to regolith) development, extended over the entire surface of Mars. This interpretation views most of Mars surface as active soil, colored red, as on Earth, by eons of widespread microbial activity [12].

6. Another new, potentially important new insight is the proposed $H_2O_2$-$H_2O$ life hypothesis [13], namely the possibility that the Martian life solvent, in the organisms detected by the LR may be $H_2O_2$-$H_2O$ rather than $H_2O$. Additionally, it is conjectured [1] that layers of structured $H_2O$ (probably vitreous, rather than crystalline, at the relevant temperatures) adsorbed on cytoskeletal/organel analogs may compartment any $H_2O_2$-$H_2O$ mixtures.

7. Collectively, these new findings and analyses, compiled with the LR data, strongly indicate microbial life on Mars. This development should re-focus the analysis of the Viking Mission results to working out the broadest physiological details required by the organisms in Marciana.

The analysis of the whole evidence thus constitutes a situation very different from that of only a few months ago. With the biological nomenclature of *Gillevinia straata,* the possibility of contamination of Marciana must be considered. This may have occurred in the missions over the past decades in which the sterilization procedures were abandoned in the belief that there was no life on Mars. This and other biosecurity concerns [14] must be evaluated. Also an epistemological objection that I have long posed, that Jakobia organisms cannot be proven extant by detection of their components alone, but only through the detection of their active metabolism [15], would seem to take on new significance. I have proposed a detailed approach that could enable the first determination of whether the Martian microor-ganisms are similar to our life forms or truly alien [16]. Further, comparative biological studies and the classification of extraterrestrial organisms could be accomplished with metabolism-detection experiments in which environmental and nutrient variables were studied. With the first extraterrestrial creature discovered and named, our sense of responsibility in this endeavor should be heightened.

———

## Acknowledgements:

Thanks to Drs. Meserve, Huntress, Hazen and others of the Carnegie Institution for providing this opportunity to speak and to discuss these issues, to Dr. Patricia Straat, my Co-Experimenter, to Drs. Schulze-Makuch and Houtkooper, for their discussions, and their seconding of the LR's detection of life albeit they propose the experiment soon killed the microorganisms detected. I am most deeply indebted to Dr. Mario Crocco, who authored the paper confirming the conclusion that the LR detected life, and to Dr. Mariela Szirko, his colleague.

I thank Argentine Minister Marcelo Cima for attending the seminar, representing Ambassador Jose Octavio Bordón, who was on official business in Puerto Rico. Particular thanks are due Governor Jorge Telerman, Governor of the Autonomous City of Buenos Aires, for having facilitated the research work concluding that the Viking Labeled Release experiment of 1976 did detect microbial life on Mars and the naming thereof.

## References:

[1] Crocco, M. (2007), Los taxones mayores de la vida orgánica y la nomenclatura de la vida en Marte: primera clasificación biológica de un organismo marciano (ubicación de los agentes activos de la Misión Vikingo de 1976 en la taxonomía y sistemática biológica). Electroneurobiología 15 (2), 1-34; http://electroneubio.secyt.gov.ar/First_biological_classification_Martian_organism.pdf

[2] Crocco, M. (2007), Corrección: primera clasificación biológica de un organismo marciano, género *Gillevinia* (no *Levinia*) / Correction note: first biological classification of a Martian organism, genus *Gillevinia* (not *Levinia*). Electroneurobiologia 15 (2), 35-37. [Included in the .pdf file above].

[3] Navarro-González, R; Navarro, K. F.; de la Rosa, J., Iñiguez, E.; Molina, P.; Miranda, L. D.; Morales, P; Cienfuegos, E.; Coll, P.; Raulin, F., Amils, R. and McKay, C. P. (2006), "The limitations on organic detection in Mars-like soils by thermal volatilization-gas chromatography-MS and their implications for the Viking results", PNAS 103 (44 ), 16089-16094. http://www.pnas.org/cgi/content/full/103/44/16089

[4] Hargraves, R.B., D.W. Collinson, R.E. Arvidson and C.R. Spitzer (1977), The Viking Magnetic Properties Experiment: Primary Mission Results. J. Geophys. Res. 82, 4547.

[5] Kerr, R.A. (2004), Life or Volcanic Belching on Mars? Science 303, # 5666, 1953, 26 March.

[6] Quam, L.O., ed. (1971), Research in the Antarctic, AAAS #93, Washington, DC.

[7] Moore, H.J. et al. (1977), Surface Materials of the Viking Landing Sites. J. Geophys. Res. 82:28, 4497-4523

[8] Dartnell, L.R., L. Desorgher, J. M. Ward, and A. J. Coates (2007), Modelling the surface and subsurface Martian radiation environment: Implications for astrobiology. Geophys. Res. Lett., 34, L02207, doi:10.1029/2006GL027494

[9] Levin, G.V.; P.A. Straat, H.P.A. Van Dongen, and J.D. Miller (2004), Circadian rhythms and evidence for life on Mars. Instruments, Methods, and Missions for Astrobiology, SPIE Proceedings 5555, 35, August.

[10] Van Dongen, H., J. Miller, P. Straat and G. Levin (2005), A circadian biosignature in the Labeled Release data from Mars?" Instruments, Methods, and Missions for Astrobiology, SPIE Proceedings 5906, OC1-10, August.

[11] G. Bianciardi (2004), Nonlinear Analysis of the Viking Lander 2 Labeled Release Data. Proc. of the III European Workshop on Exo-Astrobiology on Mars: The search for Life, Madrid, Spain, 18-20 November 2003 (ESA SP-545, March).

[12] Paepe, R. (2007), "The Red Soil on Mars as a proof for water and vegetation" Geophysical Research Abstracts 9, 01794; SRef-ID: 1607-7962/gra/EGU2007-A-01794,
http://www.cosis.net/abstracts/EGU2007/01794/EGU2007-J-01794.pdf?PHPSESSID=e

[13] Houtkooper, J. M. and D. Schulze-Makuch (2007), A Possible Biogenic Origin for Hydrogen Peroxide on Mars: The Viking Results Reinterpreted. In press at Int. J of Astrobiology. A previous version in: http://arxiv.org/pdf/physics/0610093

[14] Marks, P. (2007), Technology: Keeping alien invaders at bay. New Scientist, April 28, pp. 24-25.

[15] Szirko, M. (2007), "Comentario editorial: la cuestión epistemológica en la detección de vida en Marte", Electroneurobiología 15 (1), 183-187;
http://electroneubio.secyt.gov.ar/Acerca_de_la_vida_en_Marte_Editorial.htm

[16] Levin, G. V. (2006), "Modern Myths Concerning Life on Mars", Electroneurobiología 14 (5), 3-25;
http://electroneubio.secyt.gov.ar/Gilbert_V_Levin_Life_on_Mars_Modern_Myths.htm

———

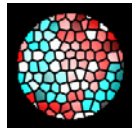